# A Cost- Effective Design of Reversible Programmable Logic Array

Pradeep Singla
M.Tech Scholer
Hindu College of Engineering
Sonipat, India

Naveen Kr. Malik
Assistant Professor
Hindu College of Engineering
Sonipat, India

## ABSTRACT
In the recent era, Reversible computing is a growing field having applications in nanotechnology, optical information processing, quantum networks etc. In this paper, the authors show the design of a cost effective reversible programmable logic array using VHDL. It is simulated on Xilinx ISE 8.2i and results are shown. The proposed reversible Programming logic array called RPLA is designed by MUX gate & Feynman gate for 3- inputs, which is able to perform any reversible 3- input logic function or Boolean function. Furthermore the quantized analysis with comparative finding is shown for the realized RPLA against the existing one. The result shows improvement in the quantum cost and total logical calculation in proposed RPLA.

## 1. INTRODUCTION
Designing of a complex digital system which dissipates low power is a competitive topic in the research field of hardware design. Because Complexity of the system will give rise to a problem of heat dissipation in the circuit and this becomes the critical limiting factor [1] [8] [9]. According to the Intel's Co-founder Gordon Moore in 1965 that the numbers of transistors on a given piece of silicon would double every couple of years. So, the complexity of the system will increases, the energy dissipation or heat dissipation will also increases with the exponential rate [9][14].

This leads to development of new computing hardware or circuits that dissipate less heat for continue to increase the computing power. In 1961, Landauer introduced that losing of bit in circuits causes the smallest amount of heat in computation and the theoretical limit of energy dissipation for losing of one bit computation is KTln2 [1]. Where K is a Boltzmann's constant equals to $1.3807 \times 10^{-23} JK^{-1}$ and T is the temperature at which the computation is performed [1]. At T=300K, this limit is $4 \times 10^{-21}$ Joules. Although the amount of energy for single bit operation in single logic of a complex circuit is decreasing ($10^{-15} J$ now a day)[12] but overall heat generation by the circuit is increasing [8] which cause the problems energy costs money, systems overheat and portable systems exhaust their batteries. Now a day's digital circuit dissipates even more energy than the theoretical [12]. Even C.H. Bennett in 1973 also showed that the dissipated energy directly correlated to the number of lost bits [2]. So, there is an alternative is to use logic operations that do not lost bits during computation. These are called reversible logic operations, and in principle they dissipate arbitrarily little heat [5] [6].

The idea of reversible computing comes from the thermodynamics which taught us the benefits of the reversible process over irreversible process. So, it is called reversible computing if its inputs can always be retrieved from its outputs [3] [4] [5] [6]. In other words, these circuits can generate unique output vector from each input vector and vice-versa. So, an N×N reversible gate can be represented as [7]

$$I_v = (I_1, I_2, I_3, \ldots\ldots\ldots\ldots\ldots, I_N)$$
$$O_v = (O_1, O_2, O_3, \ldots\ldots\ldots\ldots\ldots, O_N)$$

Where $I_v$ = input vectors

$O_v$ = output vectors

Reversible logics have very useful applications in everyday life like Laptop computers, wearable computers, spacecraft, smart cards etc [8]. So, in a new paradigm in computer design, reversible logics play a very important role over irreversible logics. The non-reversible PLA structure may be used to realize the reversible functions [9].Programming logic devices (PLDs) or Arrays (PLAs) are used in industrial applications to synthesis cost effective solutions to industrial designs [4]. The PLA designed using reversible gates is called RPLA. By seeing the benefits of the PLA structure over the non-PLA structure, we implement the new cost-effective RPLA by using MUX gate [10] & Feynman gate [5] [11]. This RPLA can be designed on single chip and by virtue of which larger circuits can be designed easily by RPLA's [8] [9]. There is an existing structure of the RPLA, proposed by *Himanshu Thapliyal et al.* which was demonstrated by using two gates i.e. Feynman gate and Fredkin gate. We propose the new structure of the RPLA which is cost effective than previous one in terms of Quantum cost and the total logical calculation performed by the design for a desired function.

This paper is present on following perspective: section II describes the background of the reversible logic and the conditions for the reversibility. Different reversible gate structure is also discussed in the same section. In Section III we propose the RPLA with its planes and in discussed their implementation. In section IV, the simulated results and the quantized parameters of the RPLA and also showing how proposed RPLA is better than existing one. In the last Section or in the Section V, we end the paper with conclusion.

## 2. REVERSIBLE LOGIC

The logical reversibility means there should be same number of output lines as the number of input lines i.e. the number of input lines and output line must be same or there should be one to one mapping between the input and output [12]. The later one i.e. physical reversibility means towards the retrieval of input by the input. The gate must be run forward and backward i.e. the input can also be recovered or retrieved from the output. When the device obeys these two conditions then the second law of thermodynamics guarantees that it dissipates no heat [12].

For logical reversibility in the digital logics there are two conditions as follows [13].

 Fan-Out is not permitted

 Feedback is not permitted

### 2.1 Review of Reversible gate structures

In the last years, several numbers of reversible gates has been proposed like Feynman gate (FG) [5], Toffolli gate (TG) [3], Fredkin gate (FRG) [4], Peres gate (PG) [6], New gate (NG, MKG, HNG and TSG, MG) [10] [11]. Here we are reviewing the Fredkin gate, Feynman gate and MG (MUX gate) [10] because these gates with different configuration are used in the paper. The combination of Fredkin gate and Feynman gate has been used in existing RPLA, but we uses the MUX gate and Feynman gate for the propose design. These two gates are best suits for the designing of as AND, OR, NOT and as data buffer which are used to implement the RPLA.

### 2.1.1 Fredkin Gate

Fig.1 shows the 3×3 reversible gate called Fredkin gate [4]. It has three inputs (A, B, C) and three outputs (P, Q, and R). The outputs are defined by P=A, Q=A'B XOR AC, R= A'C XOR AB. Quantum cost of a fredkin gate is 5.

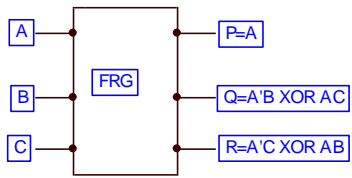

Fig.1: 3×3 Fredkin gate structure

### 2.1.2 Feynman Gate

Fig.2 shows the 2×2 reversible gate called Feynman gate [5]. Feynman gate is also recognized as controlled- not gate (CNOT). It has two inputs (A, B) and two outputs (P, Q). The outputs are defined by P=A, Q=A XOR B .This gate can be used to copy a signal. Since fan-out is not allowed in reversible logic circuits, the Feynman gate is used as the fan-out gate to copy a signal. Quantum cost of a Feynman gate is 1.

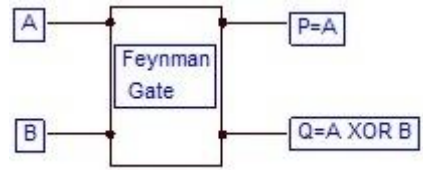

Fig.2: 2×2Feynman gate structure

#### 2.1.2.1 Feynman Gate as a Data Copier and as NOT gate

The structure of Feynman gate as Data Copier & as NOT gate is shown in the fig 2.1 & 2.2 resp. If we provide '0' at second input B then the output Q will provide the copy of first input and if we If we provide '1' at second input B then the output Q will provide the complement of the first input.

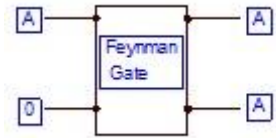

Fig 2.1: Feynman gate as Data copier

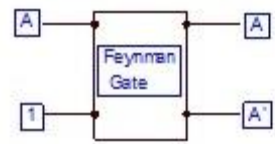

Fig.2.2: Feynman gate as NOT gate

### 2.1.3 MUX Gate

Fig.3 shows the pictorial representation of 3×3 reversible gate called MUX (MG) gate [10]. It is a conservative gate having three inputs (A, B, C) and three outputs (P, Q, R). The outputs are defined by P=A, Q=A XOR B XOR C and R= A'C XOR AB. The hamming weight of its input vector is same as the hamming weight of its output vector and its Quantum cost is 4.

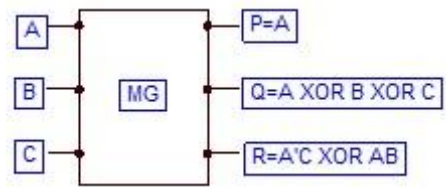

Fig.4: 3×3 MUX gate

#### 2.1.3.1 MUX Gate as AND Gate and OR Gate

The structure of MUX gate as AND gate and OR gate are shown in the fig4.1 & 4.2. If we provide '0' at third input C then the output R will provide the AND combination of first

& second input and if we If we provide '1' at second input B then the output Q will provide the OR combination of the first & third input.

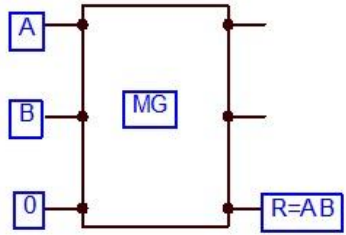

Fig4.1: MUX gate as AND gate

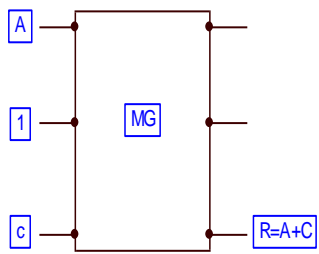

Fig.4.2: MUX gate as OR gate

## 2.2 Parameters

In conventional digital circuits, the complexity of the system i.e. number of gates is the parameter for the analysis [7]. But, in the reversible gates, along with the complexity, there are some other parameters which are used for analysis of reversible logic are as:-

*Constant Input (CO):* The number of inputs that are kept constant (0 or 1) for synthesis the given functions [15].

Quantum Cost (QC): The number of reversible gates (1×1 or 2×2) to realize the circuit is known as quantum cost [15].

Garbage Output (GO): The number of outputs that are not primary is known as Garbage outputs [15].

Total logical calculation (T): Total logical calculation is the count of the XOR, AND, NOT logic in the output circuit [15].

## 3. RPLA BLOCK IMPLEMENTATION

In this paper, we proposed the architecture of cost- effective Reversible PLA (RPLA) which does not lost information or dissipates less heat (Ideally no heat). The Block diagram of RPLA is shown in fig5

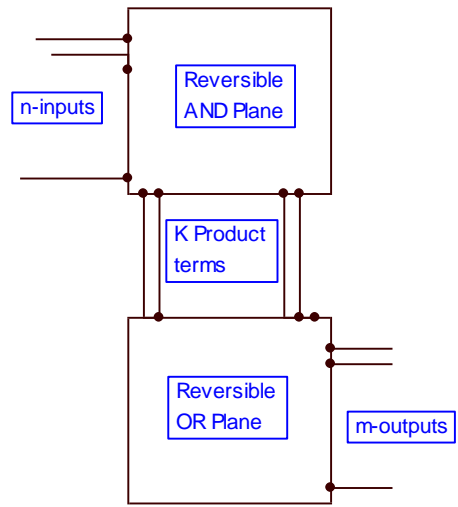

**Fig5: RPLA Block representation**

In this architecture, we used the two planes i.e. Reversible AND plane & Reversible OR plane as in conventional design. In this there are n number of inputs and m number of outputs and k number of product terms. The reversible AND plane is designed by the MUX & Feynman gates to produce the required product terms and to prevent of fan-out .The reversible AND plane is designed by the MUXgate..

For the implementation purpose we use the VHDL codes. VHDL stands for very high speed integrated circuit hardware language having three kinds of modeling i.e. Behavioral modeling, structural modeling and data flow modeling [16]. In our proposed design, first we code MUX gate, Feynman gate define using behavioral modeling and then for complete plane we connect different entities by structural modeling.

### 3.1 VHDL Codes for MUX Gate

```
Library IEEE;
Use ieee.std_logic_1164.all;
Use ieee.numeric_std.all;
Entity MG3 is
Port (IN1: in STD_LOGIC;
 IN2: in STD_LOGIC;
 IN3: in STD_LOGIC;
 OUT1: out STD_LOGIC;
 OUT2: out STD_LOGIC;
 OUT3: out STD_LOGIC);
End MG3;
Architecture Behavioral of MG3 is
Begin
OUT1<= IN1;
OUT2<=IN1 xor IN2 xor IN3;
OUT3<= (((NOT IN1) and IN3) xor (IN1 and IN2));
End behavioral;
```

### 3.2 VHDL Codes for Feynman Gate

```
library ieee;
use ieee.std_logic_1164.all;
use ieee.numeric_std.all;
```

```
entity FY2 is
port(IN1:in STD_LOGIC;

IN2:in STD_LOGIC;
OUT1:out STD_LOGIC;
OUT2:out STD_LOGIC);
end FY2;
architecture behavior of FY2 is
begin
OUT1<=IN1;
OUT2<=IN1 xor IN2;
End behavior;
```

### 3.3 Design of Reversible AND Plane
The design of reversible AND Plane of proposed RPLA is shown in fig. 9.

### 3.4 Design of Reversible OR Plane
The design of reversible OR Plane of proposed RPLA is shown in fig. 8.

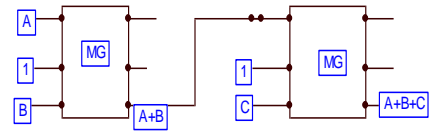

Fig.8 Block Diagram of OR Plane of RPLA

## 4. RESULTS
### 4.1 Simulation Results
The propose RPLA (AND plane and OR Plane) shown in section-3 are implemented and resulted by using VHDL and simulated in Xilinx ISE8.2i. The simulated results of the AND plane and OR plane can be seen in fig.8 and fig.9 resp.

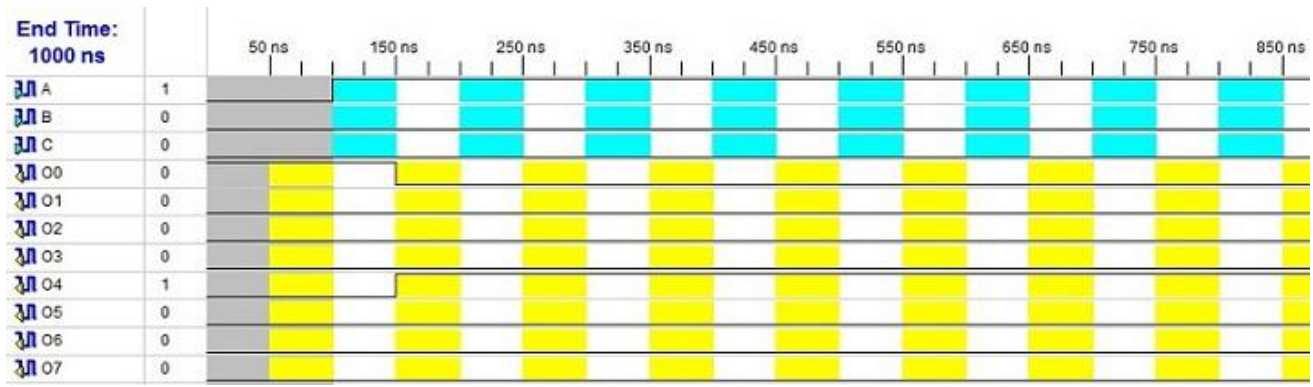

Fig.6 Simulated result of reversible AND plane

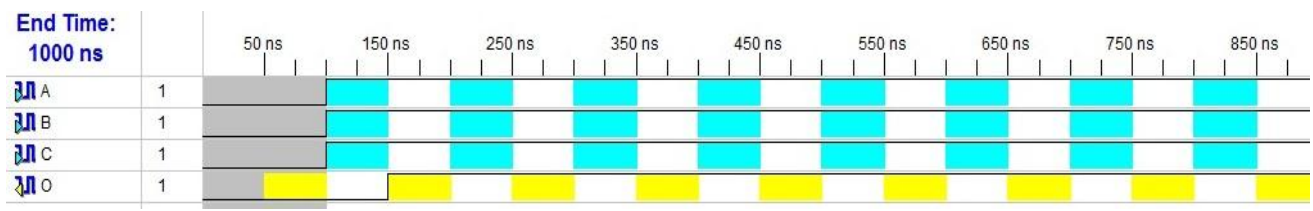

**Fig7 . Simulated result of reversible OR plane**

### 4.2 Quantized evaluation of parameters in proposed RPLA implementation
In case of reversible logic the important parameters are total logical calculations, No. of gate used, Garbage outputs, constant inputs and Quantum cost. These are calculated for existing RPLA[9] and for proposed RPLA. The comparative findings are shown in table 1.

### 4.2.1 Total Logical calculation (T)
Assuming

$\alpha$ = A two input XOR gate calculation

$\beta$ = A two input AND gate calculation

$\delta$ = A NOT gate calculation

T = Total logical calculation

The Total logical calculation is the count of the XOR, AND, NOT logic in the output circuit. For example MUX gate has three XOR gate and two AND gate and one NOT gate in the output expression. Therefore $T_{(M)} = 3\alpha + 2\beta + \delta$.

In the previous work [9], the total number of logical calculation T is: For AND Plane T= 16× (2α+4β+2δ)( for Fredkin gate)+ 21×1α(for Feynman gate) = 53α + 64β+32δ. For OR Plane T = 3(2α+4β+2δ)= 6α+12β+6δ and in our proposed implementation, , the total logical calculation for reversible AND plane is: T = 16× (3α+2β+δ)( for MUX gate)+ 21×1α(for Feynman gate) = 69α + 32β+16δ. The total logical calculation for reversible OR plane is: 3(3α+2β+δ) = 9α+6β+3δ.

### 4.2.2 Number of Reversible Gates Used
For reversible AND Plane: 37
For reversible OR Plane: 3

### 4.2.3 Garbage Outputs
In this RPLA implementation, we can see that there are two outputs of the Feynman gate corresponding to the inputs. In this the complementary function and the copying function is generated and they further utilized in the circuit. Thus, we can say that the garbage output is zero for this implementation.

### 4.2.4 Constant Inputs
The number of inputs that are kept constant ( 0 or 1) for synthesis the given functions and in this implementation we used 37 reversible gates( for and array) and 3 reversible gates ( for OR array) . All the reversible gates used constant inputs (0 or 1) for producing the different outputs. So, there are 37 CI (for AND plane) and 3 CI (for OR plane).

### 4.2.5 Quantum Cost
As we have used Mux gate whose quantum cost is 4[6]. So, if there are n number of gates used in the structure then complete quantum cost becomes 4n. In the existing RPLA, author used fredkin gate whose quantum cost is 5. So, our proposed structure for RPLA is efficient than existing one.
From the Quantized evaluations all parameters comparison with existing structure [5] are   shown in table 1.

**Table 1. Parameters comparison**

| Parameters | | Proposed | Existing[5] |
|---|---|---|---|
| Number of Gates | AND Plane | 37 | 37 |
|  | OR Plane | 3 | 3 |
| GO | AND Plane | 0 | 0 |
|  | OR Plane | 0 | 0 |
| CI | AND Plane | 37 | 3 |
|  | OR Plane | 37 | 3 |
| QC |  | f+4m | f+5$f_r$ |
| Parameters |  | Proposed | Existing[5] |
| Total Logical Calculation | AND Plane | 69α+32β+16δ | 53α+64β+32δ |
|  | OR Plane | 9α+6β+3δ | 6α+12β+6δ |

Where f = No. of Feynman gates.
   m = No. of Mux gates
   $f_r$ = No. of fredkin gates

So, from the above discussed parameters, it can be says that the number of total calculation & QC are reduced so the less hardware complexity in the proposed design than existing[9].

## 5. CONCLUSION
In this paper, we emphasis on an efficient approach to design low power digital systems using proposed RPLA. We proposed an improved design of RPLA in the paper and proving the concept of using MUX gate & Feynman gate for the design of RPLA is efficient and cost-effective than the existing one. The comparison in terms of number of gates, GO, CI and Total logical calculation has been shown in table 1 and from there it is crystal clear that the proposed design has optimized design in comparison with existing. The simulated results are also shown. So. this proposed RPLA will provide a new approach to the arena of low power reconfigurable computing hardware.

field programmable gate arrays"978-114244-8728-8/11/IEEE 2011

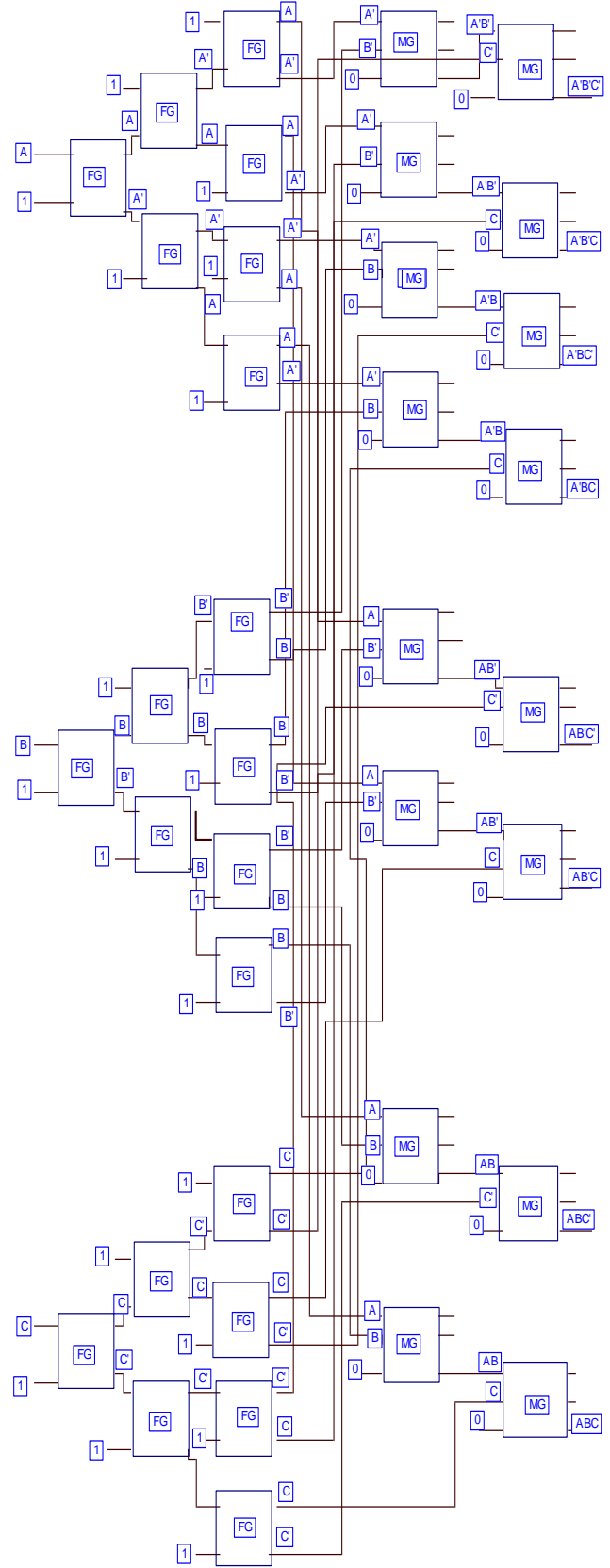

**Fig.9 Block Diagram of AND Plane of RPLA**